\documentclass[sigconf,nonacm]{acmart}
\pdfoutput=1

\acmConference[womENcourage 2021]{ACM Celebration of Women in Computing: womENcourage™ }{September 22-24}{Online}
\usepackage{enumitem}

\usepackage{fancyhdr}
\begin{document}

\title{Identifying the Prevalence of Gender Biases among the Computing Organizations}

\author{Sayma Sultana, London Ariel Cavaletto, Amiangshu Bosu}
\email{{sayma, london.cavaletto, amiangshu.bosu }@wayne.edu}
\affiliation{%
  \institution{Wayne State University}
  \city{Detroit}
  \state{Michigan}
  \country{USA}
}


\begin{abstract}
We have designed an  online survey \textit{to understand the status quo of four dimensions of gender biases among the  contemporary computing organizations.} 
Our preliminary results found almost one-third of the respondents have reported first-hand experiences of encountering gender biases at their jobs. 
\end{abstract}




\maketitle
\lhead{ACM womENcourage, September 22-24, 2021, 2021, Prague, Czech Republic}
\rhead{Sayma et al.}

\section{Motivation \& Objectives}
\label{sec:introduction}
While 37\%  of the computing workforce were women in 1995, that number is expected to drop to 21\% by 2022\footnote{\url{https://girlswhocode.com/about-us}}. The contemporary computing industry not only lacks gender diversity but also is not gender inclusive, as the majority of the women computing professionals are concentrated in the lower echelon. The diversity and inclusion crisis among the computing organizations can be partially blamed on various biases that women in computing are often subject to. For example, in STEM-related professions, around 50\% of women experience gender inequality or sexual harassment \cite{team_dynamics}. Women in computing also get less recognition for their accomplishments than men, and often are dissatisfied with their career trajectory. As a result, 45\% of women leave tech jobs within ten years and the turn over rate is more than twice as high for women than it is for men~\cite{ashcraft2016women}. A recent survey among the professionals who voluntarily left their tech jobs found that LGBTQ employees were the most likely to experience public humiliation (24\%) and to be bullied (20\%) than non-LGBTQ employees(13\%)\cite{nbcnews}. Harassment and underpayment hinder them to be productive and connect with people at their workplaces. To improve the retention of women and LGBTQ personas in computing professions,  it is crucial to identify and eradicate biases against them.  While prior studies have identified various facets of biases against women in computing, the prevalence of such biases against female and LGBTQ people, and their consequences need further investigations. Therefore, we have designed a study \textit{to understand the status quo of gender biases among the  computing organizations.} 
In this study, we focus on following four dimensions of gender biases, which have been frequently reported by prior studies.

\begin{enumerate}[leftmargin=*]

\item \textbf{ Lack of career development opportunities: } A common saying goes, a woman has to be 2X as good as men to get 0.5X recognition as her male contemporaries. Often, deserving women are not considered for leadership positions and career progression of female software engineers is throttled before they are promoted to management roles. Madeline et al. demonstrate that motherhood impedes female employees from reaching for traditional male positions\cite{motherhood}, the problem which is coined as \textit{maternal wall}.

\item \textbf{ Project or task selection:} Not only hostile but also benevolent norms foster biased attitudes in the workplace\cite{kuchynka}. Whereas hostile work norm involves assuming women incompetent or talking over them in formal meetups, benevolent work practice includes assigning women more mundane or simpler assignments, providing extra help with challenging tasks. Both of these norms turn down the sense of belonging for female employees and negatively impact their job performance.
   
 \item \textbf{Unwanted sexual attention:} Acts considered as harassment to women may seem normal gendered interaction or fun to men\cite{quinn}. Also, men tend to exhibit sexist behaviour more when their masculine identity is threatened. 
 
 
\item \textbf{Harassment using misogynistic humour or gestures:} This form involves harassing people based on their sexual orientation or preferences. When people come out as LGBTQ, they become the target of exclusive attitudes or humours that denigrate or derogate their identities. Many organizations lack policies to make LGBTQ people feel safe in their workplaces and overlook problems of this group while attempting to build gender inclusive workplaces\cite{LGBTQ_ref}.

\end{enumerate}

To identify the frequency and impact of biased decisions or attitudes among the computing organizations, we designed an anonymous online survey and distributed it to computing professionals worldwide. We received 78 complete responses till now. As of May 2021, response collection for our survey is ongoing. 


\section{Research Method}
In the survey, we focus on collecting both first-hand and second-hand experiences of contemporary computing professionals regarding the four categories of gender biases described in the Section 1. Our survey employs the scenario-based questionnaire design, which has been commonly used to investigate biased interactions. We wrote four scenarios with each scenario focusing on one of the four gender bias dimensions. Each scenario in our survey: i) is relatively short (within 150 words), ii) includes at least two personas, where one persona represents a woman or LGBTQ+ person as the victim and another persona as the perpetrator, and iii) interactions between the personas demonstrate a biased behavior. 
We created the scenarios based on a  prior study investigating gender biases\cite{kuchynka}, developer blogs,  and stories shared on  devRant\footnote{https://devrant.com/}, an online community for software developers. In the following, we provide a brief overview of the four scenarios.

\begin{enumerate}[leftmargin=*]

    \item The first scenario focuses on `lack of career development opportunities' for the persona Melissa, who is a hard-working female. Her boss Sean (M) selects Asish, another male, for a promotion to a managerial position despite of Melissa being both senior and more qualified  than Asish. Sean thinks that Melissa's motherhood will impact her performance.
    
    \item  In the `biased task assignment' scenario, a male persona named Eric, assigns low complexity tasks e.g. documentation, frontend development to Gianna, a female and new hire in the company. When Gianna encounters problems, instead of helping her to solve it, Eric passes that task to Steve, who is a male and hired at the same time as Gianna.
    
    \item In the `unwanted sexual attention scenario', the persona named Shruti, who is a female, receives excessive attention from Kevin (a male), who  works on the same project as Shruti. Despite of being turned down by Shruti multiple times, Kevin keeps asking her out. When Shruti unofficially complains about Kevin to her supervisor,  Steve (a male), he laughs it away.
    
    \item In the `crude or sexist jokes' scenario, the persona named Mikehl is a gay male software developer. Mikehl brings his partner, Christian, to company's anniversary party and introduces  Christian to his colleagues. After that  Salvatore ( a straight male) starts making crude remarks  targeting the sexual orientation of Mikehl during conversations as well as in code reviews.
\end{enumerate}

In the survey, after presenting each scenario, we asked the respondents if a similar scenario occurred in their workplaces. If they answered in positive, we divided the remaining questions into three conditional group of questions: 1. If they experienced this form of bias as recipient 2. If they witnessed or heard about such experience. 3. If they created such scenario for others.  Each group of questions had both closed and open-ended questions about the frequency and the effect of the occurrences. This survey has been approved by Wayne State University’s Institutional Review Board (IRB) and required 15 minutes on average to be completed. We shared the survey through social media and emails to invite software engineers to participate in this study. We focused on reaching out to minorities since they are more likely to encounter bias in workplaces .




\section{ Preliminary Results and Ongoing works}
\label{sec:implication}

 \begin{figure}
	  \includegraphics[width=\linewidth]{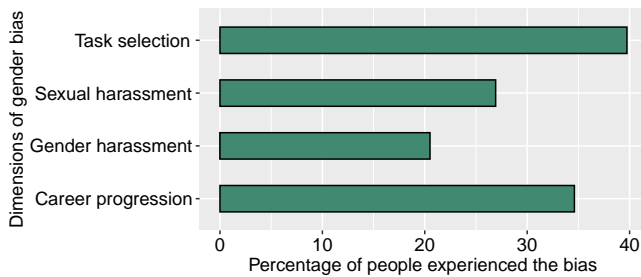}
	\caption{Percentage of respondents reporting gender biases }
	\label{fig:building-expertise}	
	\vspace{-12pt}
\end{figure}

We  have received 78 complete responses so far from this ongoing study. Among the respondents,  38 (48.7\%) are females, 35 (44.9\%) are males and remaining 4 ( 5.1\%) are nonbinaries. Figure 1 shares the preliminary result of the survey. As it shows,  35\%  and 31\% of our respondents have encountered biases regarding career progression and task selection in workplace respectively. That means one among every three people experienced some sort of gender bias in software industry. One respondent shared her frustration about career development: \textit{"I was doing all of the things my manager and senior manager were telling me I needed to do to "prove" I was worthy, and it didn't matter at all... I kept thinking I must be doing something wrong. I must be missing something."}
Another respondent mentioned about changing her career trajectory due to bias regarding task selection: \textit{"I have been, and will always be, angry to be considered a technical person of a lesser god.  I hold an engineering degree from one of the most prestigious institutions in the US, and have moved away from coding into executive leadership, but I am constantly challenged: "Is she technical enough?" As if I found my degree in a CrackerJack box."}

Respondents also shared that they required higher standard for getting promotion, were rejected for pay-raise despite having worthy accomplishments, had been treated unfairly when they became mothers. As a result, they got frustrated and lost interest in their job. Furthermore, even being in leading positions, female engineers encounter \textit{mansplaining} for the same idea they have shared, assigned with trivial tasks. Often women do not realize such attitudes as bias against them assuming they are not trying hard enough and give rise to self doubts. 

The preliminary  findings of this study align with prior studies and support the prevalence of gender bias in software industry. We also show how bias negatively impacts female engineers' performance and job satisfaction. Though companies are taking initiatives to ensure diversity in workplaces, inclusion still remains out of focus. Knowledge about the frequency and impact of biased incidents in software industry can raise awareness and assist people to make neutral decisions in workplaces.

We plan to invite more software developers to fill out this survey. We are particularly focusing on reaching out to communities of women or LGBTQ+ computing professionals. We hope that at the completion of this study, we will able to provide a more clear quantification  these four categories of biases among contemporary computing organizations.





\bibliographystyle{ACM-Reference-Format}
\bibliography{security-references}


\begin{thebibliography}{7}


\ifx \showCODEN    \undefined \def \showCODEN     #1{\unskip}     \fi
\ifx \showDOI      \undefined \def \showDOI       #1{#1}\fi
\ifx \showISBNx    \undefined \def \showISBNx     #1{\unskip}     \fi
\ifx \showISBNxiii \undefined \def \showISBNxiii  #1{\unskip}     \fi
\ifx \showISSN     \undefined \def \showISSN      #1{\unskip}     \fi
\ifx \showLCCN     \undefined \def \showLCCN      #1{\unskip}     \fi
\ifx \shownote     \undefined \def \shownote      #1{#1}          \fi
\ifx \showarticletitle \undefined \def \showarticletitle #1{#1}   \fi
\ifx \showURL      \undefined \def \showURL       {\relax}        \fi
\providecommand\bibfield[2]{#2}
\providecommand\bibinfo[2]{#2}
\providecommand\natexlab[1]{#1}
\providecommand\showeprint[2][]{arXiv:#2}

\bibitem[\protect\citeauthoryear{Ashcraft, McLain, and Eger}{Ashcraft
  et~al\mbox{.}}{2016}]%
        {ashcraft2016women}
\bibfield{author}{\bibinfo{person}{Catherine Ashcraft}, \bibinfo{person}{Brad
  McLain}, {and} \bibinfo{person}{Elizabeth Eger}.}
  \bibinfo{year}{2016}\natexlab{}.
\newblock \bibinfo{booktitle}{\emph{Women in tech: The facts}}.
\newblock \bibinfo{publisher}{National Center for Women \& Technology (NCWIT)}.
\newblock


\bibitem[\protect\citeauthoryear{Brammer}{Brammer}{2017}]%
        {nbcnews}
\bibfield{author}{\bibinfo{person}{John~Paul Brammer}.} \bibinfo{year}{April
  29, 2017}\natexlab{}.
\newblock \bibinfo{title}{Bullying Is Driving LGBTQ People Out of Tech, Study
  Finds}.
\newblock
  \bibinfo{howpublished}{\url{https://www.nbcnews.com/feature/nbc-out/bullying-driving-lgbtq-people-out-tech-study-finds-n752646}}.
\newblock
\newblock
\shownote{[Online; accessed June 20, 2021].}


\bibitem[\protect\citeauthoryear{Heilman and Okimoto}{Heilman and
  Okimoto}{2008}]%
        {motherhood}
\bibfield{author}{\bibinfo{person}{Madeline Heilman} {and}
  \bibinfo{person}{Tyler Okimoto}.} \bibinfo{year}{2008}\natexlab{}.
\newblock \showarticletitle{Motherhood: A Potential Source of Bias in
  Employment Decisions}.
\newblock \bibinfo{journal}{\emph{The Journal of applied psychology}}
  \bibinfo{volume}{93} (\bibinfo{date}{02} \bibinfo{year}{2008}),
  \bibinfo{pages}{189--98}.
\newblock
\urldef\tempurl%
\url{https://doi.org/10.1037/0021-9010.93.1.189}
\showDOI{\tempurl}


\bibitem[\protect\citeauthoryear{James, Galster, Blincoe, and Miller}{James
  et~al\mbox{.}}{2017}]%
        {team_dynamics}
\bibfield{author}{\bibinfo{person}{Toni James}, \bibinfo{person}{Matthias
  Galster}, \bibinfo{person}{Kelly Blincoe}, {and} \bibinfo{person}{Grant
  Miller}.} \bibinfo{year}{2017}\natexlab{}.
\newblock \showarticletitle{What Is the Perception of Female and Male Software
  Professionals on Performance, Team Dynamics and Job Satisfaction? Insights
  from the Trenches}. \bibinfo{pages}{13--22}.
\newblock
\urldef\tempurl%
\url{https://doi.org/10.1109/ICSE-SEIP.2017.31}
\showDOI{\tempurl}


\bibitem[\protect\citeauthoryear{Kuchynka, Bosson, Vandello, and
  Puryear}{Kuchynka et~al\mbox{.}}{2018}]%
        {kuchynka}
\bibfield{author}{\bibinfo{person}{Sophie Kuchynka}, \bibinfo{person}{Jennifer
  Bosson}, \bibinfo{person}{Joseph Vandello}, {and} \bibinfo{person}{Curtis
  Puryear}.} \bibinfo{year}{2018}\natexlab{}.
\newblock \showarticletitle{Zero‐Sum Thinking and the Masculinity Contest:
  Perceived Intergroup Competition and Workplace Gender Bias}.
\newblock \bibinfo{journal}{\emph{Journal of Social Issues}}
  \bibinfo{volume}{74} (\bibinfo{date}{09} \bibinfo{year}{2018}),
  \bibinfo{pages}{529--550}.
\newblock
\urldef\tempurl%
\url{https://doi.org/10.1111/josi.12281}
\showDOI{\tempurl}


\bibitem[\protect\citeauthoryear{Quinn}{Quinn}{2002}]%
        {quinn}
\bibfield{author}{\bibinfo{person}{Beth~A. Quinn}.}
  \bibinfo{year}{2002}\natexlab{}.
\newblock \showarticletitle{Sexual Harassment and Masculinity: The Power and
  Meaning of ‘Girl Watching}.
\newblock \bibinfo{journal}{\emph{Gender \& Society 16}}  \bibinfo{volume}{3}
  (\bibinfo{date}{06} \bibinfo{year}{2002}), \bibinfo{pages}{386–402}.
\newblock
\urldef\tempurl%
\url{https://doi.org/10.1177/0891243202016003007.}
\showDOI{\tempurl}


\bibitem[\protect\citeauthoryear{Talves}{Talves}{2016}]%
        {LGBTQ_ref}
\bibfield{author}{\bibinfo{person}{Kairi Talves}.}
  \bibinfo{year}{2016}\natexlab{}.
\newblock \showarticletitle{Discursive self-positioning strategies of Estonian
  female scientists in terms of academic career and excellence}.
\newblock \bibinfo{journal}{\emph{Women's Studies International Forum}}
  \bibinfo{volume}{54} (\bibinfo{year}{2016}), \bibinfo{pages}{157--166}.
\newblock
\showISSN{0277-5395}
\urldef\tempurl%
\url{https://doi.org/10.1016/j.wsif.2015.06.007}
\showDOI{\tempurl}


\end{thebibliography}


\end{document}